\documentclass{webofc}
\usepackage[varg]{txfonts}  
\begin{document}

\title{Probing Massive Star Nucleosynthesis with Data on Metal-Poor Stars and the Solar System}

\author{\firstname{Yong-Zhong} \lastname{Qian}\inst{1}\fnsep\thanks{\email{qianx007@umn.edu}}
}

\institute{School of Physics and Astronomy, University of Minnesota, Minneapolis, Minnesota 55455, USA
}

\abstract{
Metal-poor stars were formed during the early epochs when only massive stars had time to evolve and contribute 
to the chemical enrichment. Low-mass metal-poor stars survive until the present and provide fossil records of 
the nucleosynthesis of early massive stars. On the other hand, short-lived radionuclides (SLRs) in the early solar 
system (ESS) reflect the nucleosynthesis of sources that occurred close to the proto-solar cloud in both space and time. 
Both the ubiquity of Sr and Ba and the diversity of heavy-element abundance patterns observed in single metal-poor stars
suggest that some neutron-capture mechanisms other than the $r$-process might have operated in early massive stars.
Three such mechanisms are discussed: the weak $s$-process in non-rotating models with initial carbon enhancement,
a new $s$-process induced by rapid rotation in models with normal initial composition, and neutron-capture processes
induced by proton ingestion in non-rotating models. In addition, meteoritic data are discussed to constrain the 
core-collapse supernova (CCSN) that might have triggered the formation of the solar system and provided some of 
the SLRs in the ESS. If there was a CCSN trigger, the data point to a low-mass CCSN as the most likely
candidate. An $11.8\,M_\odot$ CCSN trigger is discussed. Its nucleosynthesis, the evolution of its remnant, and 
the interaction of the remnant with the proto-solar cloud appear to satisfy the meteoritic constraints and can 
account for the abundances of the SLRs $^{41}$Ca, $^{53}$Mn, and $^{60}$Fe in the ESS.
}

\maketitle

\section{Data on Metal-Poor Stars}
\label{mpdata}
Observations show that heavy elements such as Sr (with mass numbers $A\sim 86$--88) and Ba ($A\sim 134$--138)
are prevalent in metal-poor stars \cite{roderer}. At the early times when these stars were formed, low-mass stars would
not have had time to evolve and produce such elements through the slow ($s$) neutron-capture process. 
Massive stars evolve quickly to end in core-collapse supernovae (CCSNe) and the neutrino-driven winds from the 
proto-neutron stars created in these events can produce Sr. Typical winds, however, do not have the appropriate conditions 
to make Ba (e.g., \cite{qian1,hoffman}). While the rapid ($r$) neutron-capture process in rare CCSNe (e.g., \cite{fischer})
or neutron star mergers (e.g., \cite{thielemann}) can produce Ba, it is difficult for such rare events to account for the
wide-ranging enrichment at early times as indicated by the ubiquity of Ba in metal-poor stars (e.g., \cite{qian2,argast}).
Therefore, some other neutron-capture processes associated with massive stars are required.

Observations also show diverse abundance patterns of heavy elements in metal-poor stars. For example, there are
substantial deviations of the abundance ratio La/Eu ($A=139$ for La and $A=151$, 153 for Eu) from the solar
$r$-process value for stars with metallicities as low as $[{\rm Fe/H}]=\log({\rm Fe/H})-\log({\rm Fe/H})_\odot\sim -2.6$ 
\cite{simmerer}. This diversity also calls for massive stellar sources for neutron-capture processes other than the 
$r$-process.

All of the above observations concern
single metal-poor stars. So both the ubiquity of Ba and the diversity of La/Eu in these stars reflect enrichment of the 
general interstellar medium (ISM) by massive stellar sources, not some localized surface contamination through
mass transfer of e.g., the $s$-process products from binary companions.

\section{Neutron-Capture Processes in Early Massive Stars}
\label{ncap}
We focus on three types of neutron-capture processes in early massive stars to address the above observations.

\subsection{Enhanced Weak $s$-Process}
\label{weaks}
Observations (e.g., \cite{aoki}) show that there exist carbon-enhanced metal-poor (CEMP) stars with [C/Fe]~$\geq 0.7$, 
some of which do not have any enhancements of neutron-capture elements. These so-called CEMP-no stars constitute 
$\sim 20\%$, 40\%, and 80\% of all the metal-poor stars with [Fe/H]~$\leq -2$, $-3$, and $-4$, respectively. 
Such stars were most likely formed out of the general ISM that had been enriched by very early CCSNe with elevated 
abundances of C (and presumably also N and O) relative to Fe. 
While such stars have low masses in order to survive until the present, they must have been formed along with massive 
CEMP stars that already exploded as CCSNe.

During core H burning of a massive star, the initial CNO nuclei are converted into $^{14}$N,
which produces $^{22}$Ne through $^{14}{\rm N}(\alpha,\gamma)^{18}{\rm F}(e^+\nu_e)^{18}{\rm O}(\alpha,\gamma)^{22}{\rm Ne}$
during core He burning. The activation of $^{22}{\rm Ne}(\alpha,n)^{25}{\rm Mg}$ during core He burning and subsequent evolution
provides neutrons for the weak $s$-process that produces heavy elements typically up to Sr. For massive CEMP stars with
initial abundances of ${\rm [CNO/H]}\gtrsim-0.5$, even heavier elements up to Ba can be produced \cite{banerjee2018a}. 
In general, the efficiency of the $s$-process in massive CEMP stars is sensitive to the initial enhancement of CNO 
and mass of the star while the yield increases approximately linearly with the initial Fe abundance. These results were obtained 
for non-rotating stars, and therefore, differ from similar results for fast-rotating massive metal-poor stars without C enhancement
(e.g., \cite{pignatari,frischknecht}) in that the former are independent of the uncertain rotation-induced mixing processes.
The enhanced weak $s$-process in non-rotating CEMP stars of $\gtrsim 20\,M_\odot$ with ${\rm [CNO/H]}\gtrsim -1.5$ can be 
an important source for heavy elements in the early Galaxy \cite{banerjee2018a}. 

\subsection{New $s$-Process Induced by Rapid Rotation}
\label{news}
Massive metal-poor stars rotating above a critical speed can reach the so-called quasi-chemically homogeneous 
(QCH) state following core H burning (e.g., \cite{yoon,woosley}). Rotation-induced mixing results in primary production of 
$^{13}$C. The subsequent occurrence of $^{13}{\rm C}(\alpha,n)^{16}{\rm O}$ during core He burning provides neutrons
for a prolific $s$-process. Depending on the rotation speed and the mass loss rate, elements up to Bi ($A=209$) can be produced
for progenitors with initial metallicities of ${\rm [Fe/H]}\lesssim -1.5$ \cite{banerjee2019}. This model of the $s$-process
in rotating massive metal-poor stars differs from other studies of the $s$-process in such stars (e.g., \cite{frischknecht})
because the QCH state was not reached for the latter. 

The above model suggests that rapidly-rotating massive metal-poor stars are likely the first sites of the main $s$-process, 
which is usually associated with slowly-evolving low-mass stars. With the above massive stellar sources, the observed 
early onset of the $s$-process can be explained. In addition, the $s$-process contributions of these sources can account 
for at least some of the CEMP-$s$ and CEMP-$r/s$ stars with strong enrichments that are attributed to the $s$-process 
and a mixture of the $r$-process and the $s$-process, respectively \cite{banerjee2019}.

\subsection{Neutron-Capture Processes Induced by Proton Ingestion}
\label{pingest}
During the last few years of the life of a metal-free or metal-poor star of $\sim 20$--$30\,M_\odot$, the He shell becomes 
convective following the depletion of C at the center. Some of the protons present at low levels in the outer He shell 
can be ingested into the inner He shell by convective boundary mixing. Subsequent transport to the hotter region initiates 
the reaction sequence $^{12}{\rm C}(p,\gamma)^{13}{\rm N}(e^+\nu_e)^{13}{\rm C}(\alpha,n)^{16}{\rm O}$, which provides 
neutron densities appropriate for driving a nuclear flow intermediate ($i$) between the $s$-process and the $r$-process, 
the so-called $i$-process \cite{cowan}. As a result, elements up to Bi are produced \cite{banerjee2018b}.
Depending on the time available before core collapse, $^{17}{\rm O}(\alpha,n)^{20}{\rm Ne}$ may occur to provide much 
lower neutron densities typical of the $s$-process to facilitate further neutron capture on the products of the preceding 
$i$-process. Consequently, the final yield pattern can vary from $s$-like to $r/s$-like \cite{banerjee2018b}.

The above model of neutron-capture processes induced by proton ingestion in early non-rotating massive stars
have important implications for the abundances of heavy elements in metal-poor stars. First of all, it provides
a rapidly-evolving source that occurred frequently to enrich the early ISM with heavy neutron-capture elements.
Therefore, it can account for the observed ubiquity of Ba in metal-poor stars (see Sect.~\ref{mpdata}). Further,
because its yield pattern can vary from $s$-like to $r/s$-like, it can also explain the diversity of heavy-element 
abundance patterns in metal-poor stars (see Sect.~\ref{mpdata}). Finally, mass transfer of $s$-process and
$i$-process products from an intermediate-mass binary companion is commonly invoked to account for CEMP-$s$ 
and CEMP-$r/s$ stars, respectively. For some CEMP-$s$ and CEMP-$r/s$ stars, however, there are no clear 
indications that they are in binaries. In fact, some CEMP-$s$ stars are observed to be single \cite{hansen}, which
is difficult for the binary mass transfer scenario to explain. Single CEMP-$s$ and CEMP-$r/s$ stars, however,
can be accounted for by the above model because they could have been formed directly from the early ISM 
that had already been enriched by massive stars with neutron-capture processes induced by proton ingestion.

\section{CCSN Contributions to the Solar System}
\label{ess}
In contrast to metal-poor stars formed at early times, both low-mass and massive stars contributed to the 
elemental abundances in the solar system, which was formed $\sim 9$~Gyr after the big bang. 
The number of CCSNe among these contributing sources can be estimated as follows. A CCSN can enrich
$\sim 3\times 10^4\,M_\odot$ of ISM (e.g., \cite{thornton}). For a Galactic CCSN rate of $\sim(30\ {\rm yr})^{-1}$
associated with $\sim 10^{10}\,M_\odot$ of gas, an average ISM would receive CCSN contributions once
every $\sim 10$~Myr. So the ISM out of which the solar system was formed, the proto-solar cloud, received 
contributions from $\sim 900$ CCSNe over $\sim 9$~Gyr.

Many short-lived radionuclides (SLRs) with lifetimes of $\sim 0.1$--10 Myr are produced by CCSNe.
Based on the above estimated interval of $\sim 10$~Myr between successive CCSN contributions,
the last few CCSNe contributing to the abundances in the solar system might have provided a significant
amount of SLRs to the early solar system (ESS). Any SLRs from those CCSNe would have been incorporated 
into the meteorites formed in the ESS. Studies of such meteoritic samples indeed found a number of SLRs
(e.g., \cite{davis}). These meteoritic data provide a strong constraint on the scenario that the last contributing
CCSN also triggered the formation of the solar system because the SLRs contributed by that CCSN 
cannot exceed the corresponding abundances in the ESS. 

For a fraction $f$ of the ejecta from the triggering CCSN injected into the proto-solar cloud 
and a time $\Delta$ between the CCSN and incorporation of SLRs into the ESS solids, 
the resulting number ratio of an SLR ($R$) to its stable isotope ($I$) would be
\begin{equation}
\left(\frac{N_R}{N_I}\right)_{\rm ESS}=\frac{fY_R/A_R}{X_{I,\odot}M_\odot/A_I}\exp\left(-\frac{\Delta}{\tau_R}\right),
\label{nri}
\end{equation}
where $Y_R$ is the CCSN yield of $R$, $\tau_R$ and $A_R$ are its lifetime and mass number, respectively,
and $X_{I,\odot}$ and $A_I$ are the solar abundance and mass number of $I$, respectively.
The necessary condition for the triggering CCSN is that $(N_R/N_I)_{\rm ESS}$ match the data on some SLRs
without exceeding the data on all the other SLRs.

An additional meteoritic constraint on the triggering CCSN comes from the lack of large deviations in the number 
ratios of stable isotopes of such major elements as Mg, Si, Ca, and Fe \cite{jerry}. Both this constraint and the
necessary condition regarding SLRs render it difficult for a high-mass CCSN to be the trigger because
the large yields of such a source would have exceeded the data on some SLRs and caused large deviations 
in the number ratios of stable isotopes of some major elements \cite{banerjee2016}. Consequently, if a CCSN 
triggered the formation of the solar system, it was most likely at the lower end of the relevant mass range 
\cite{banerjee2016,andre}.

\section{Low-Mass CCSN Trigger for Solar System Formation}
\label{ccsn}
The nucleosynthesis of an $11.8\,M_\odot$ CCSN was studied in \cite{banerjee2016,andre}. 
Using a 3D model of the explosion, \cite{andre} found that the SLRs $^{41}$Ca, $^{53}$Mn, and $^{60}$Fe 
in the ESS can be accounted for by such a CCSN (see Table~\ref{slr}) with $f=3.6\times10^{-6}$ and
$\Delta=0.675$~Myr [see Eq.~(\ref{nri})]. Because only a very small fraction of the CCSN ejecta needs
to have been injected into the proto-solar cloud, the contributions to the other SLRs and the stable
isotopes were negligible.

\begin{table}
\centering
\caption{Example results on SLRs in the ESS.}
\label{slr}
\begin{tabular}{lccccr}
\hline
$R/I$ & $\tau_R$ (Myr) & $Y_R\ (M_\odot)$ & $X_{I,\odot}$ &  $(N_R/N_I)_{\rm ESS}$ & Data\\\hline
$^{41}{\rm Ca}/^{40}{\rm Ca}$ & 0.147 & $7.11\times10^{-6}$ & $5.88\times10^{-5}$ & $4.31\times10^{-9}$ & 
$(4.6\pm1.9)\times10^{-9}$\\
$^{53}{\rm Mn}/^{55}{\rm Mn}$ & 5.40 & $2.73\times10^{-5}$ & $1.29\times10^{-5}$ & $6.97\times10^{-6}$ & 
$(7\pm1)\times10^{-6}$\\
$^{60}{\rm Fe}/^{56}{\rm Fe}$ & 3.78 & $4.17\times10^{-6}$ & $1.12\times10^{-3}$ & $1.04\times10^{-8}$ & 
$(1.01\pm0.27)\times10^{-8}$\\\hline
\end{tabular}
\end{table}

The parameters $f$ and $\Delta$ for the above low-mass CCSN trigger are consistent with the evolution 
of a CCSN remnant in an ISM and with the interaction between the associated shock and the proto-solar cloud.
Simulations showed that a fraction $\epsilon_{\rm in}\sim 0.05$ of the CCSN ejecta impacting the cloud
would be injected into the cloud \cite{boss2014,boss2015}. The fraction $f$ of the net CCSN ejecta injected into 
a cloud of radius $r_c\sim 0.1$~pc \cite{bergin} can be estimated as
\begin{equation}
f\sim\epsilon_{\rm in}\left(\frac{\delta\Omega_c}{4\pi}\right)\sim\frac{\epsilon_{\rm in}}{4}\left(\frac{r_c}{R_s}\right)^2
\sim 1.25\times10^{-4}\left(\frac{\epsilon_{\rm in}}{0.05}\right)\left(\frac{r_c}{0.1\ {\rm pc}}\right)^2\left(\frac{\rm pc}{R_s}\right)^2,
\end{equation}
where $\delta\Omega_c$ is the solid angle subtended by the cloud from the center of the CCSN and $R_s$ is the 
shock radius upon impact. To obtain $f\sim3.6\times10^{-6}$ requires $R_s\sim 5.9$~pc. In addition, to trigger
the collapse of the cloud requires a shock velocity $v_s\sim20$--40~km~s$^{-1}$ \cite{boss2014,boss2015}.

The evolution of $R_s$ and $v_s$ for a CCSN remnant \cite{cioffi} can be estimated as
\begin{eqnarray}
R_s&=&14.0\left(\frac{E}{10^{51}\ {\rm erg}}\right)^{2/7}\left(\frac{{\rm cm}^{-3}}{n_0}\right)^{3/7}
\left(\frac{4}{3}t_*-\frac{1}{3}\right)^{3/10}\ {\rm pc},\\
v_s&=&413\left(\frac{E}{10^{51}\ {\rm erg}}\right)^{1/14}\left(\frac{n_0}{{\rm cm}^{-3}}\right)^{1/7}
\left(\frac{4}{3}t_*-\frac{1}{3}\right)^{-7/10}\ {\rm km\ s}^{-1},
\end{eqnarray}
where $E$ is the CCSN explosion energy, $n_0$ is the hydrogen number density of the ISM, and 
$t_*=t/t_{\rm unit}$ is the time $t$ since the explosion in units of
\begin{equation}
t_{\rm unit}=1.33\times10^4\left(\frac{E}{10^{51}\ {\rm erg}}\right)^{3/14}\left(\frac{{\rm cm}^{-3}}{n_0}\right)^{4/7}\ {\rm yr}.
\end{equation}
For $E=2\times 10^{50}$~erg \cite{muller} and $n_0\sim 50$~cm$^{-3}$ \cite{bergin},
$R_s\sim 5.9$~pc is reached for $t\sim5.3\times10^4$~yr with an impact velocity $v_s\sim 33$~km~s$^{-1}$.
Because the above evolution time is far less than $\Delta$, the latter would be essentially
the time needed for condensation of solids in the ESS.

\section{Discussion and Conclusions}
\label{concl}
Observations of single metal-poor stars reveal both the ubiquity of Sr and Ba and the diversity of heavy-element 
abundance patterns, which appear to require some neutron-capture mechanisms other than the $r$-process to
have operated in early massive stars. Among various possibilities, perhaps the weak $s$-process in non-rotating 
massive CEMP stars is the best understood. Its contributions might have been significant at low metallicities 
if a substantial fraction of the stars of $\gtrsim 20\,M_\odot$ had been formed with ${\rm [CNO/H]}\gtrsim -1.5$
\cite{banerjee2018a}. The new $s$-process induced by rapid rotation can potentially account for the observed 
early onset of the $s$-process and at least some of the CEMP-$s$ and CEMP-$r/s$ stars \cite{banerjee2019}.
The modeling of this process depends on our understanding of the evolution of rapidly-rotating metal-poor stars 
and the associated mixing. Likewise, the neutron-capture processes induced by proton ingestion in 
early non-rotating massive stars depend on the treatment of mixing during the evolution of such stars. These 
processes can potentially account for both the observed ubiquity of Ba and the diversity of heavy-element 
abundance patterns in metal-poor stars \cite{banerjee2018b}. In addition, their relatively frequent contributions
to the ISM may have resulted in the formation of a significant fraction of the CEMP-$s$ and CEMP-$r/s$ stars 
by birth, which could have been the dominant channel to form such stars in single configuration 
\cite{banerjee2018b}.

Meteoric data demonstrate the presence of SLRs in the ESS. They also show a lack of large deviations in the 
number ratios of stable isotopes of such major elements as Mg, Si, Ca, and Fe. These data render it difficult 
for a high-mass CCSN to be the trigger for the formation of the solar system because the large yields of such
a source would have exceeded the data on some SLRs and caused large deviations in the number ratios of 
stable isotopes of some major elements \cite{banerjee2016}. Consequently, if there was a CCSN trigger, 
a low-mass CCSN would be the most likely candidate \cite{banerjee2016,andre}.
The nucleosynthesis of an $11.8\,M_\odot$ CCSN, the evolution of its remnant, and the interaction of
the remnant with the proto-solar cloud appear to satisfy the meteoritic constraints and can account for the 
abundances of the SLRs $^{41}$Ca, $^{53}$Mn, and $^{60}$Fe in the ESS.

\section*{Acknowledgments}
I thank Projjwal Banerjee, Wick Haxton, Alexander Heger, Bernhard M\"uller, and Andre Sieverding for
collaboration. This work was supported in part by the U.S. Department of Energy under Grant No.
DE-FG02-87ER40328.

\end{document}